\documentclass[12 pt,twoside,aps,prd,amsmath,amssymb,
tightenlines,showpacs,showkeys,eqsecnum]{revtex4-1}

\usepackage{graphicx}
\usepackage{mathptmx}
\usepackage{bm}
{\hspace*{\fill}{\protect\small Victor Rikhvitsky, B. Saha, M.
Visinescu} \hspace*{\fill} }
{\hspace*{\fill} {\protect\small {\bf Magnetic Bianchi type II
string cosmological model in loop quantum cosmology}}
\hspace*{\fill} } 

\def\w{\mathring{\omega}}
\def\q{\mathring{q}}
\def\e{\mathring{e}}

\def\myfigure #1#2#3#4
{\begin{figure}[ht]\begin{center}
\includegraphics[width=#2 \textwidth]{#1.eps}
\parbox[t]{#4\textwidth}{\caption{#3}\label{#1}}
\end{center}\end{figure}}

\def \myfigures #1#2#3#4#5#6#7#8
{\begin{figure}[ht]
    \begin{center}
        \includegraphics[width=#2 \textwidth]{#1.eps}
        \hfill
        \includegraphics[width=#6 \textwidth]{#5.eps}
        \parbox[t]{#4\textwidth}{\caption {#3}\label{#1}}
        \hfill
        \parbox[t]{#8\textwidth}{\caption {#7}\label{#5}}
    \end{center}
\end{figure} }

\begin{document}
\baselineskip -24pt
\title{Magnetic Bianchi type II string cosmological model in loop
quantum cosmology}
\author{Victor Rikhvitsky}\email{rqvtsk@jinr.ru}
\affiliation{Laboratory of Information Technologies\\
Joint Institute for Nuclear Research\\
141980 Dubna, Moscow region, Russia}

\author{Bijan Saha}\email{bijan@jinr.ru}
\affiliation{Laboratory of Information Technologies\\
Joint Institute for Nuclear Research\\
141980 Dubna, Moscow region, Russia}

\author{Mihai Visinescu}\email{mvisin@theory.nipne.ro}
\affiliation{Department of Theoretical Physics\\
National Institute for Physics and Nuclear Engineering\\
Magurele, P. O. Box MG-6, RO-077125 Bucharest, Romania}

\vskip 1 cm

\begin{abstract}

The loop quantum cosmology of the Bianchi type II string cosmological
model in the presence of a homogeneous magnetic field is studied.
We present the effective  equations which provide modifications to
the classical equations of motion due to quantum effects. The numerical
simulations confirm that the big bang singularity is resolved by
quantum gravity effects.

\end{abstract}

\keywords{Bianchi type II  model, cosmological string, magnetic
field, loop quantum cosmology}

\pacs{04.60.-m; 04.60.Pp; 98.80.Qc}

\maketitle

\bigskip

%%%%%%%%%%%%%%%%%%%%%%%%%%%%%%%%%%%%%%%%%%%%%%%%%%%%%%%%%%%%%%%%%%%%%
\section{Introduction}

Loop quantum cosmology (LQC) arises from the application of the more
general theory of loop quantum gravity (LQG) to cosmology. One of the
most important predictions of LQC is that in the homogeneous and
isotropic Friedmann-Robertson-Walker  models the classical
big bang singularity is avoided being replaced by a bouncing Universe.

More recently, it has been shown that the big bang singularity is also
resolved for anisotropic Bianchi type I (BI) \cite{AWEI}, II (BII)
\cite{AWEII} and IX  \cite{WEIX}. Usually the matter source that
was considered is a massless scalar field that play the role of internal
time. The investigations have been extended to more complicated models
including a perfect fluid, magnetic fields \cite{MV}, cosmological
strings \cite{RSV}. It is remarkable the fact that in all these studies
it was observed that the bounce prediction is robust.

The purpose of this paper is to investigate the dynamics of a BII
string cosmological model in the presence of a magnetic field in the
framework of LQC. We show that a bounce occurs in a collapsing
magnetized BII string cosmological model, thus extending the known
cases of singularity resolution.

The plan of the paper is as follows: In Sect. 2 we outline the
classical equations of a BII string cosmological model in the presence
of a magnetic field. In Sect. 3 we discuss the quantum theory introducing
the effective equations for the model. Sect. 4 is devoted to numerical
calculations and it is shown that the classical singularity is resolved
in the BII effective quantum dynamics. Finally, in Sect. 5 we summarize
our results.

%%%%%%%%%%%%%%%%%%%%%%%%%%%%%%%%%%%%%%%%%%%%%%%%%%%%%%%%%%%%%%%%%%
\section{Classical dynamics}
%%%%%%%%%%%%%%%%%%%%%%%%%%%%%%%%%%%%%%%%%%%%%%%%%%%%%%%%%%%%%%%%
\subsection{Hamiltonian formulation}

The spacetime metric of BII model is given by
\begin{equation}
ds^{2} = - N(t)^2 dt^2 + a_1(t)^2(dx - K zdy)^2 + a_2(t)^2 dy^2
+  a_3(t)^2 dz^2,
\label{BII}
\end{equation}
where $a_1,\,a_2$ and $a_3$ are the directional scale factors.
The parameter $K$ makes the difference between BI ($K=0$) and BII
($K=1$) spacetimes.

Having in view a comparison between the classical theory and the
effective theory from LQC it is useful to rewrite the theory in
terms of triads and connections.

Taking into account that the spatial manifold is non-compact, one
needs to introduce a fiducial cell ${\mathcal V}$ with the coordinate
lengths $l_i$ and the fiducial volume $V_0 = l_1 l_2 l_3$ \cite{AWEII}.
The fiducial metric is
\begin{equation}
\q_{ab} := \delta_{ij} \w_a^i \w_b^j \,,
\end{equation}
with the co-triads
\begin{equation}
\w_a^1 = (dx)_a - Kz (dy)_a \quad , \quad \w_a^2 = (dy)_a \quad , \quad
\w_a^3 = (dz)_a \,,
\end{equation}
and triads
\begin{equation}
\e_1^a = \left( \frac{\partial}{\partial x} \right)^a \quad , \quad
\e_2^a = Kz \left( \frac{\partial}{\partial x} \right)^a +
\left( \frac{\partial}{\partial y} \right)^a \quad , \quad
\e_3^a = \left( \frac{\partial}{\partial z} \right)^a \,.
\end{equation}

In terms of the fiducial triads $\e_i^a$ and co-triads $\w_a^i$ a
convenient parame\-tri\-za\-ti\-on of the phase space variables
$E^a_i\,, A^a_i$ is
\begin{equation}
A^i_a = c^i (l_i)^{-1} \w^i_a \quad , \quad E^a_i = p_i l_i V_o^{-1}
\sqrt{\q} \,,
\end{equation}
without sum over index $i$.

The connection and triad components $c^i$ and $p_i$ satisfy the
Poisson brac\-ket
\begin{equation}
\{c^i\,, p_j\} = 8 \pi G \gamma \delta^i_j\,,
\end{equation}
where $\gamma \approx$ 0.2375 is Barbero-Immirzi parameter.

Choosing the lapse function $N = \sqrt{|p_1 p_2 p_3|}$, in the
Hamiltonian formulation of the model we have the Hamiltonian constraint
\cite{AWEII,CM,GS}
\begin{equation}
\begin{split}
{\mathcal H}_{cl} = &\frac{-1}{8 \pi G \gamma^2}\biggl[ p_1 p_2 c_1 c_2+
p_2 p_3 c_2 c_3 + p_1 p_3 c_1 c_3 + K\epsilon p_2 p_3 c_1  \biggr.\\
&\biggl. - (1 + \gamma^2) \left( \frac{K p_2 p_3}{2 p_1}\right)^2
\biggr] + {\mathcal H}_M V = 0 \,,
\end{split}
\end{equation}
where
\begin{equation}\label{vol}
V = \sqrt{|p_1 p_2 p_3|}  \,,
\end{equation}
denotes the physical volume of the cell ${\mathcal V}$.
$\epsilon = \pm 1$ depending on whether the frame $\e^a_i$ is right or
left handed. Without any loss of generality, we will choose the
orientation to be positive in the following.
The Hamiltonian for the matter  contribution ${\mathcal H}_M$
is proportional to the matter energy density
\begin{equation}
{\mathcal H}_M  = \rho_M V \,.
\end{equation}

The triads $p_i$ are related to the directional scale factors as
\begin{equation}
p_1 = l_2 l_3 a_2 a_3 \quad , \quad p_2 = l_1 l_3 a_1 a_3 \quad , \quad
p_3 = l_2 l_1 a_1 a_1 \,,
\end{equation}
assuming $a_i >0$, i.e. $p_i>0$ with the positive orientation of the
triads.

The relation between the phase space variables $c_i$ and the metric
variables are \cite{CM}
\begin{subequations}
\begin{eqnarray}
c_1 &= \gamma l_1 a_1 H_1 + \frac{K}{2} \frac{a_1^2 l_1^2}{a_2 a_3
l_2 l_3} \,,\\
c_2 &= \gamma l_2 a_2 H_2 - \frac{K}{2} \frac{a_1 l_1}{a_3 l_3} \,,\\
c_3 &= \gamma l_3 a_3 H_3 - \frac{K}{2} \frac{a_1 l_1}{a_2 l_2} \,,
\end{eqnarray}
\end{subequations}
in terms of the Hubble parameters
\begin{equation}\label{Hub}
H_i = \frac{\dot{a_i}}{a_i} \quad, \quad i =1,2,3\,,
\end{equation}
where the 'dot' represents the derivative with respect to the harmonic
time.
%%%%%%%%%%%%%%%%%%%%%%%%%%%%%%%%%%%%%%%%%%%%%%%%%%%%%%%%%%%%%%
\subsection{Einstein's equation}

Einstein's equations are derived from Hamilton's equations:
\begin{equation}\label{ham}
\dot{p_i}= \{p_i,\mathcal{H}_{cl}\} =  -\kappa \gamma \frac{\partial
\mathcal{H}_{cl}}{\partial c_i} \quad ,\quad \dot{c_i}=
\{c_i,\mathcal{H}_{cl}\} =  \kappa \gamma \frac{\partial
\mathcal{H}_{cl}}{\partial p_i}\,,
\end{equation}
where $\kappa = 8 \pi G$.

Using the explicit form of the Hamiltonian $\mathcal{H}_{cl}$ we have
the following equations \cite{AWEII,CM}:

\begin{subequations}
\label{hamp}
\begin{eqnarray}
\dot{p_1} &= \frac{1}{\gamma} (p_1 p_2 c_2 + p_1 p_3 c_3 +
K p_2 p_3)\,,\\
\dot{p_2} &= \frac{1}{\gamma} (p_2 p_1 c_1 + p_2 p_3 c_3)
\,,\\
\dot{p_3} &= \frac{1}{\gamma} (p_3 p_1 c_1 + p_3 p_2 c_2)
\,,
\end{eqnarray}
\end{subequations}

\begin{subequations}
\label{hamc}
\begin{eqnarray}
\dot{c_1} &= -\frac{1}{\gamma} \left(p_2 c_1 c_2  + p_3 c_1 c_3 +
\frac{1}{2 p_1} (1+ \gamma^2)\left(\frac{K p_2 p_3}
{p_1}\right)^2\right) \nonumber\\
&+ \kappa \gamma p_2 p_3 \left (\rho_M + p_1 \frac{\partial \rho_M}
{\partial p_1}\right )\,,\\
\dot{c_2} &= -\frac{1}{\gamma} \left(p_1 c_2 c_1  +
p_3 c_2 c_3 + K p_3 c_1
- \frac{1}{2 p_2} (1+ \gamma^2)\left(\frac{K p_2 p_3}
{p_1}\right)^2\right)\nonumber\\
&+ \kappa \gamma p_1 p_3 \left (\rho_M + p_2 \frac{\partial \rho_M}
{\partial p_2}\right )\,,\\
\dot{c_3} &= -\frac{1}{\gamma} \left(p_1 c_3 c_1  +
p_2 c_3 c_2 + K  p_2 c_1
-\frac{1}{2 p_3} (1+ \gamma^2)\left(\frac{K p_2 p_3}
{p_1}\right)^2\right)\nonumber\\
&+ \kappa \gamma p_1 p_2 \left (\rho_M + p_3 \frac{\partial \rho_M}
{\partial p_3}\right )\,.
\end{eqnarray}
\end{subequations}

From the above equations of motion it can be observed that the
classical solutions posses the following constants of motion:

\begin{subequations}
\label{const}
\begin{eqnarray}
c_1 p_1 + c_2 p_2 &:= C_{12}\,,\\
c_1 p_1 + c_3 p_3 &:= C_{13}\,,\\
c_3 p_3 - c_2 p_2 &= C_{32} = C_{13} - C_{12} \,,
\end{eqnarray}
\end{subequations}
with $C_{12}\,, C_{13}$ constants. These equations allow us to find
exact analytically solutions for $p_1$ and $p_2$:

\begin{subequations}
\label{p2p3}
\begin{eqnarray}
\dot{p}_2 = \gamma^{-1} p_2 (c_1 p_1 + c_3 p_3) = \gamma^{-1} p_2
C_{13} \quad & \Rightarrow \quad p_2 = p_2^0
\exp\left(\frac{C_{13} \, t}{\gamma}\right)\,, \\
\dot{p}_3 = \gamma^{-1} p_3 (c_1 p_1 + c_2 p_2) =
\gamma^{-1} p_3 C_{12} \quad & \Rightarrow \quad p_3 = p_3^0
\exp\left(\frac{C_{12} \, t}{\gamma}\right)\,,
\end{eqnarray}
\end{subequations}
with $p_2^0 \,, p_3^0$ the initial values at $t=0$ of these
variables.
%%%%%%%%%%%%%%%%%%%%%%%%%%%%%%%%%%%%%%%%%%%%%%%%%%%%%%
\subsection{Cosmic strings in the presence of a magnetic field}

In our model the matter density $\rho_M$ comprises the contribution
of cosmological string density $\rho_{string}$ and the energy density
of the magnetic field
\begin{equation}\label{rhoM}
\rho_M = \rho_{string} + \rho_{mag}\,.
\end{equation}

The energy momentum tensor for a system of cosmic strings and magnetic
field in a comoving coordinate system is given by
\begin{equation}
T_\mu^\nu = \rho_{string} u_\mu u^\nu - \lambda x_\mu x^\nu +
E_\mu^\nu \,,
\end{equation}
where $\rho_{string}$ is the rest energy density of strings with
massive particles attached to them \cite{PSL}. It can be expressed as
\begin{equation}
 \rho_{string} = \rho_{p} + \lambda \,,
\end{equation}
where $\rho_{p}$ is the rest energy of the particles attached to the
strings and $\lambda$ is the tension density of the system of strings.
The four velocities $u_i$ and the direction of the string $x_i$ obey
the relations
\begin{equation}
u_\mu u^\mu  = -x_\mu x^\mu  = -1, \quad u_\mu  x^\mu  = 0\,.
\end{equation}

The electromagnetic field $E_{\mu\nu}$ is taken in the form given by
Lichnerowich \cite{lich}. We assume that the magnetic field is
homogeneous aligned
along the $z$-direction, $B_\mu \sim B_3 \delta ^3_\mu$ and consequently
the energy density of the magnetic field is \cite{SRV,MV}
\begin{equation}
\rho_{mag} = \frac{{\mathcal J}^2}{2 \mu (a_1 a_2)^2}\,,
\end{equation}
where ${\mathcal J}$ is a constant and $\mu$ is the magnetic
permeability of the medium. Typically $\mu$ differs from unity
only by a few parts in $10^{-5}$ and in our numerical simulations we
shall take $\mu = 1$.

Taking into account the conservation of the energy-momentum tensor,
i.e., $T_{\mu;\nu}^{\nu} = 0$, after a little manipulation one obtains
\cite{SV,SRV,RSV}:
\begin{equation}\label{rholambda}
\dot \rho_{string} + \frac{\dot V}{V}\rho_{string} - \frac{\dot
a_1}{a_1}\lambda = 0\,.
\end{equation}
Here we take into account that the conservation law for magnetic
field fulfills identically.

Usually it is assumed that $\rho_{string}$ and $\lambda$ are
proportional \cite{PSL}:
\begin{equation}\label{rhoalphalambda}
\rho_{string} = \alpha \lambda \,,
\end{equation}
where the constant $\alpha$ is $1$ for the so called geometric string,
greater than $1$ for Takabayasi string and $-1$ for Reddy string.

The solution of  \eqref{rholambda} with the proportionality relation
\eqref{rhoalphalambda} is
\begin{equation}\label{rhostring}
\rho_{string} = R a_1^{\frac{1-\alpha}{\alpha}} a_2^{-1} a_3^{-1}\,,
\end{equation}
with $R$ a constant of integration.

%%%%%%%%%%%%%%%%%%%%%%%%%%%%%%%%
%%%%%%%%%%%%%%%%%%%%%%%%%%%%%%%%%%%%
\section{Effective dynamics within LQC}

In LQC the connection variables $c_i$ do not have
direct quantum analogues and are replaced by holonomies.
The quantum effects are incorporated in the {\it effective}
Hamiltonian, ${\mathcal H}_{eff}$,
constructed from the classical one, ${\mathcal H}_{cl}$,  by replacing
the connection components $c_i$ with sine functions:
\begin{equation}\label{csin}
c_i \longrightarrow \frac{\sin (\bar \mu_i c_i)}{\bar \mu_i}\,.
\end{equation}
where $\bar \mu_i$ are real valued functions of the triad coefficients
$p_i$.

In what follows we shall use the so called $\bar \mu'$-scheme in which
the parameters $\bar \mu_i'$ are chosen as follows \cite{CV,C}:
\begin{equation}
\bar \mu_1' = \sqrt{\frac{p_1 \Delta}{p_2 p_3}}\quad,\quad
\bar \mu_2' = \sqrt{\frac{p_2 \Delta}{p_1 p_3}}\quad,\quad
\bar \mu_3' = \sqrt{\frac{p_3 \Delta}{p_1 p_2}}\,,
\end{equation}
where $\Delta = 4\sqrt{3} \pi \gamma l_{Pl}^2  $ is the
area gap in the LQC with the Planck length
$l_{Pl} :=\sqrt{G\hbar/c^3}$.

The equations of motion \eqref{ham} that incorporate loop quantum
modifications \eqref{csin} are deduced accordingly \cite{AWEII,CM,GS}.
For the equations for $c_i$ in r.h.s. we shall take into account the
contribution of the matter density \eqref{rhoM}.

The equations  for the effective theory are given by Poisson
brackets with the Hamiltonian constraint (${\mathcal H}_{eff}=0$):
\begin{subequations}
\begin{eqnarray}
\dot{p_1} &= \frac{p_1^2}{\gamma\bar\mu_1}\left(\sin\bar\mu_2c_2+
\sin\bar\mu_3c_3+ \eta \right)\cos\bar\mu_1c_1\,, \\
 \dot{p_2} &= \frac{p_2^2}{\gamma\bar\mu_2}(\sin\bar\mu_1c_1+
 \sin\bar\mu_3c_3)\cos\bar\mu_2c_2\,, \\
 \dot{p_3} &= \frac{p_3^2}{\gamma\bar\mu_3}(\sin\bar\mu_1c_1+
 \sin\bar\mu_2c_2)\cos\bar\mu_3c_3\,,
 \end{eqnarray}
\end{subequations}
\begin{align}
\dot{c_1} &= -\frac{p_2p_3}{2\gamma\Delta}\left[\frac{}{}
2(\sin\bar\mu_1c_1\sin\bar\mu_2c_2+\sin\bar\mu_1c_1\sin\bar\mu_3c_3
+\sin\bar\mu_2c_2\sin\bar\mu_3c_3) \nonumber \right.\\
& \qquad +{\bar\mu_1c_1}\cos\bar\mu_1c_1
(\sin\bar\mu_2c_2+\sin\bar\mu_3c_3)-{\bar\mu_2c_2}\cos\bar\mu_2c_2
(\sin\bar\mu_1c_1+\sin\bar\mu_3c_3) \nonumber \\
& \qquad -{\bar\mu_3c_3}\cos\bar\mu_3c_3
(\sin\bar\mu_1c_1+\sin\bar\mu_2c_2)+ \eta^2 (1+\gamma^2) \nonumber \\
& \left. \qquad +\eta(\bar\mu_1c_1\cos\bar\mu_1c_1-\sin\bar\mu_1c_1)
\frac{}{}\right]\nonumber\\
&\quad +R\frac{\alpha-1}{2\alpha} p_1^{-\frac{1+\alpha}{2\alpha}}
p_2^{\frac{1+\alpha}{2\alpha}}p_3^{\frac{1+\alpha}{2\alpha}}
+\frac{{\cal J}^2}{2\mu} p_1 p_2 p_3^{-1}\,,
\end{align}
\begin{align}
\dot{c_2} &= -\frac{p_1p_3}{2\gamma\Delta}\left[\frac{}{}
2(\sin\bar\mu_1c_1\sin\bar\mu_2c_2+\sin\bar\mu_1c_1\sin\bar\mu_3c_3
+\sin\bar\mu_2c_2\sin\bar\mu_3c_3) \nonumber \right.\\
& \qquad -{\bar\mu_1c_1}\cos\bar\mu_1c_1
(\sin\bar\mu_2c_2+\sin\bar\mu_3c_3)+{\bar\mu_2c_2}\cos\bar\mu_2c_2
(\sin\bar\mu_1c_1+\sin\bar\mu_3c_3) \nonumber \\
& \left.\qquad -{\bar\mu_3c_3}
\cos\bar\mu_3c_3(\sin\bar\mu_1c_1+\sin\bar\mu_2c_2)\right]
-\eta^2(1+\gamma^2) \nonumber \\
& \left.\qquad - \eta(\bar\mu_1c_1\cos\bar\mu_1c_1 -3\sin\bar\mu_1c_1)
\frac{}{}\right]\nonumber\\
&\quad +R\frac{\alpha+1}{2\alpha} p_1^{\frac{\alpha-1}{2\alpha}}
p_2^{\frac{1-\alpha}{2\alpha}}p_3^{\frac{1+\alpha}{2\alpha}}
+\frac{{\cal J}^2}{2\mu} p_1 p_3^{-1}\,,
\end{align}
\begin{align}
\dot{c_3} &= -\frac{p_1p_2}{2\gamma\Delta}\left[\frac{}{}
2( \sin\bar\mu_1c_1\sin\bar\mu_2c_2+\sin\bar\mu_1c_1\sin\bar\mu_3c_3
+\sin\bar\mu_2c_2\sin\bar\mu_3c_3) \nonumber \right.\\
& \qquad -{\bar\mu_1c_1}\cos\bar\mu_1c_1
(\sin\bar\mu_2c_2+\sin\bar\mu_3c_3)-{\bar\mu_2c_2}\cos\bar\mu_2c_2
(\sin\bar\mu_1c_1+\sin\bar\mu_3c_3) \nonumber \\
& \left.\qquad +{\bar\mu_3c_3}
\cos\bar\mu_3c_3(\sin\bar\mu_1c_1+\sin\bar\mu_2c_2)\right]
-\eta^2(1+\gamma^2) \nonumber \\
& \left.\qquad -\eta(\bar\mu_1c_1\cos\bar\mu_1c_1 -3\sin\bar\mu_1c_1)
\frac{}{}\right] \nonumber\\
&\quad +R\frac{\alpha+1}{2\alpha} p_1^{\frac{\alpha-1}{2\alpha}}
p_2^{\frac{1+\alpha}{2\alpha}}p_3^{\frac{1-\alpha}{2\alpha}}
-\frac{{\cal J}^2}{2\mu} p_1 p_2 p_3^{-2}\,.
\end{align}
where $\eta = K \sqrt{\Delta p_1^{-3} p_2 p_3}$.
The complexity of these equations does not allow for analytic solutions
and imposes numerical simulations.
However, taking into account that $\sin\theta$ is bounded by $1$ for
all $\theta$ it is possible to infer an upper bound for the density
of matter \cite{AWEII}:
\begin{equation}\label{rhomax}
\rho_M \leq \frac{3+ (1 + \gamma^2)^{-1}}{8 \pi G \gamma^2 \Delta}
\approx 0.54 \, \rho_{Pl}\,.
\end{equation}
Let us note that in the case of BI cosmology, the
matter density is bounded by 0.41\,$\rho_{Pl}$.
%%%%%%%%%%%%%%%%%%%%%%%%%%%%%%%%%%%%%%%%%%%%%%%%%%%%%%%%%%%%%%%%%%%
\section{Numerical solutions}

In our numerical analysis we choose the units: $\hbar = c = G =1, \,
\gamma = 0.2375,\, l_1 = l_2 = l_3 =1,\, \epsilon =1,\, \mu =1, K=1$.
The graphics are plotted as functions of the harmonic time (lapse
function $N = \sqrt{|p_1 p_2 p_3|}$).

In order to compare the classical solutions with those of the effective
equations we investigate the evolution of the directional Hubble
parameters \eqref{Hub},
the matter density \eqref{rhoM}, the volume \eqref{vol} and the shear
\begin{equation}
\Sigma^2 = \frac{1}{6}[ (H_1 - H_2)^2 + (H_2 - H_3)^2 +
(H_1 - H_3)^2]\,.
\end{equation}

In the figures  red solid line shows the volume scale $V=a_1
a_2 a_3\,,$ blue dashed line presents the density of matter $\rho_M$,
and black dotted lines is for the shear $\Sigma^2$.

The initial conditions for effective and classical solutions should
be chosen in accordance with the Hamiltonian constraints. In order
to simplify the analysis we choose as initial conditions $c_2(t_0) =
c_3(t_0) = c_0$ and $p_1(t_0)=p_2(t_0)=p_3(t_0)=p_0$ at the initial
moment $t=t_0$. In Fig. \ref{pc} we plot the admissible values of
$c_0, p_0$ which provide an acceptable value for the initial data
$c_1(t_0)$.

First of all we choose the initial conditions corresponding to a
classically collapsing universe approaching the classical big bang
singularity. For this purpose we take all initial $c_i$ as being
negative. In Fig. \ref{4103} we have plotted the classical evolution
of volume scale $V$, matter density $\rho_{M}$ and shear $\Sigma^2$
for a positive proportionality constant  $\alpha = +1$ with
nonvanishing magnetic field (${\cal I}\neq 0$) and string  density
($R\neq 0$). The Fig. \ref{1103} shows quantum analog of Fig.
\ref{4103}. The Fig. \ref{4003} is the classical picture of
evolution with a negative proportionality constant  $\alpha = -1$.
The quantum counterpart of Fig. \ref{4003} is given in Fig.
\ref{1003}. As it is seen from the Figs. \ref{4103} - \ref{1003},
while the classical evolution ends in Big Crunch in quantum case the
spacetime avoids singularity and after attaining some minimum value
the Universe again begins to expand. It should be noted that the
qualitative pictures of evolution are almost the same for positive
or negative values of $\alpha$. Accordingly, in what follows we
shall present only the plots for a single value of the parameter
$\alpha$, namely $\alpha =+1$.

\myfigure{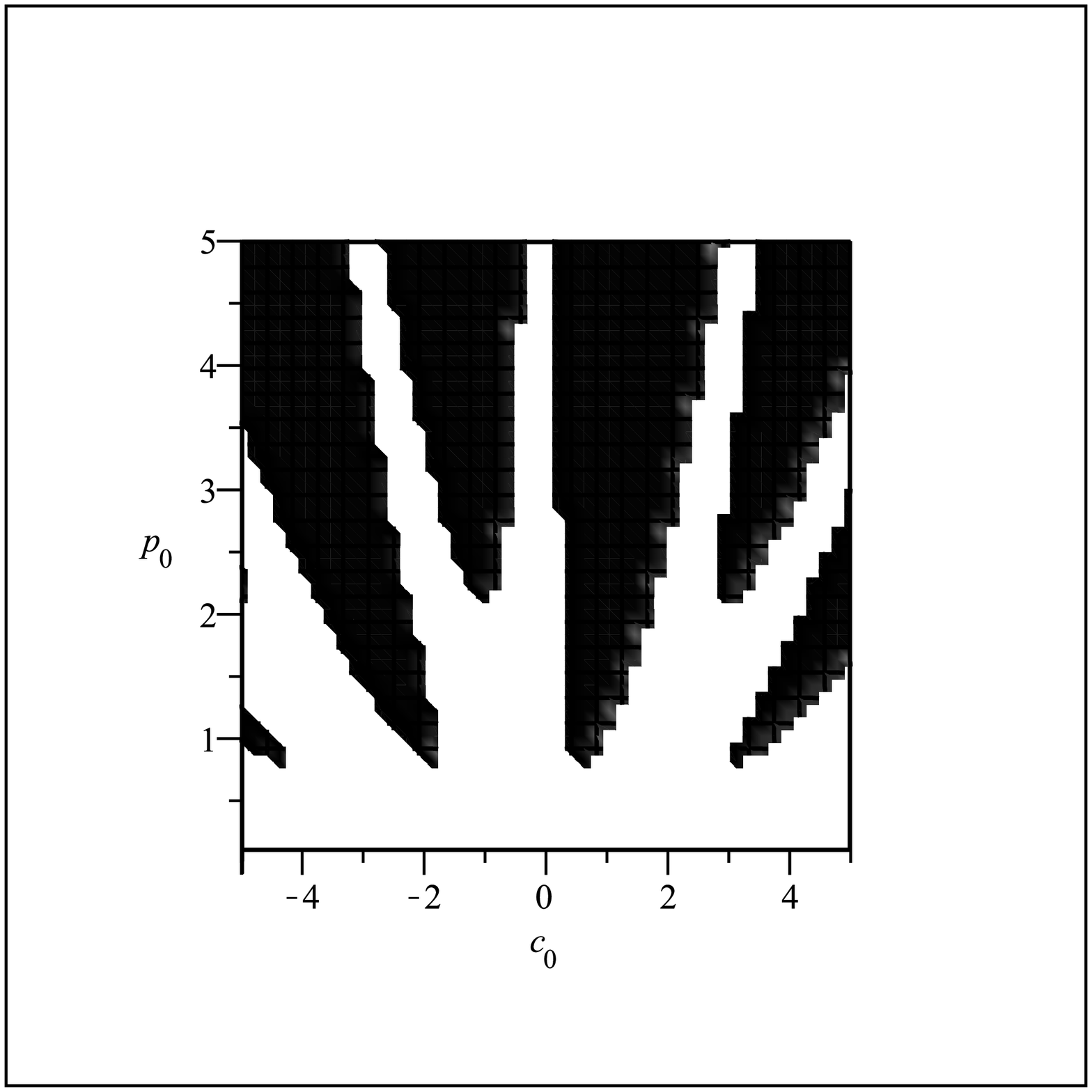}{0.75}{The sectors in which $c_0$ and $p_0$ take
admissible values defined by the constrain $\mathcal{H}_{eff} = 0$.
The black region in the plot corresponds to the admissible values of
these parameters. The figure is valid for any combinations of $R =
0$, $R = 0.2$, ${\cal I} = 0$, ${\cal I} = 0.5$.}{0.75}

\myfigures{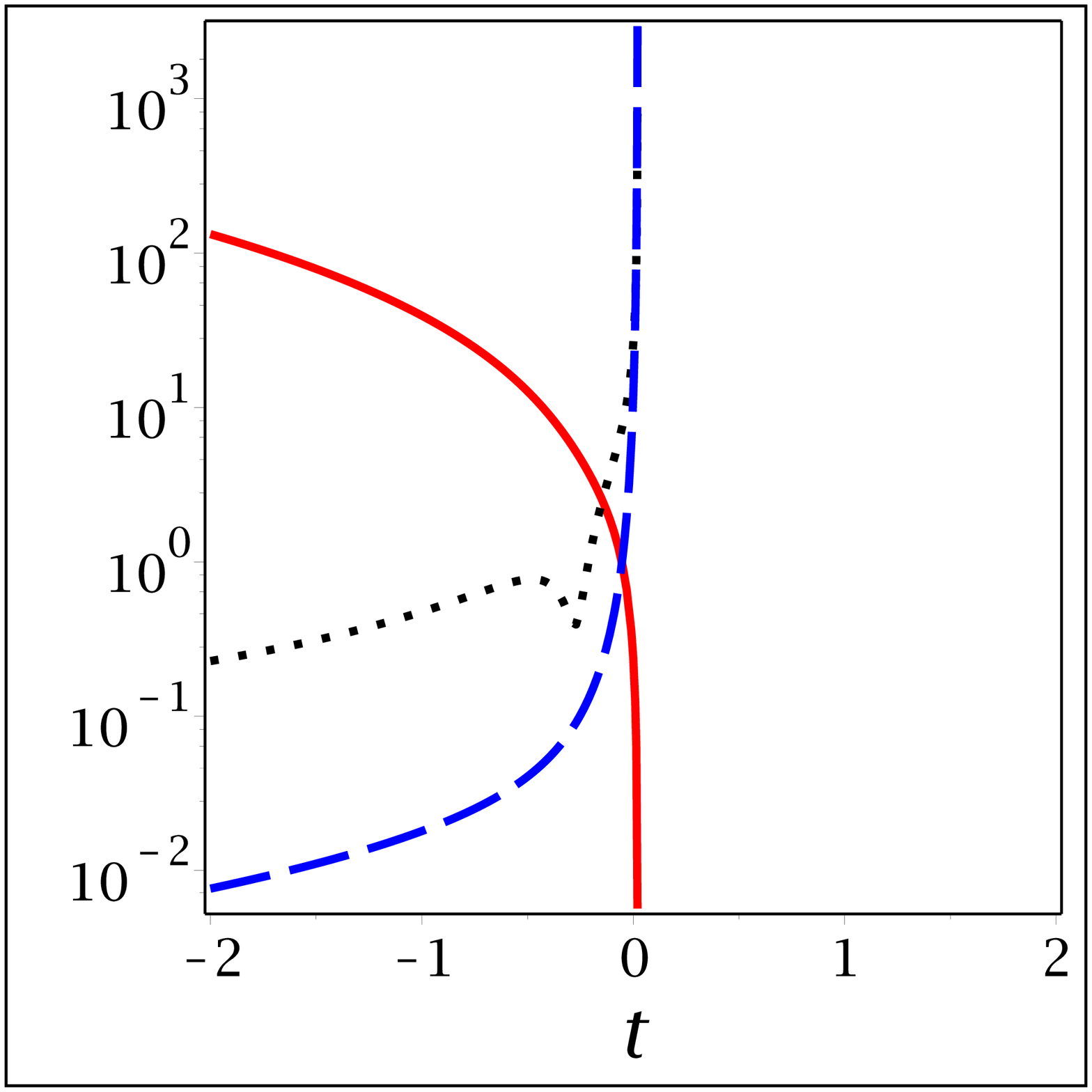}{0.45}{ Classical case for $c_0=-1$, ${\cal I}=0.5$,
$R=0.2$, $\alpha=+1$ and $c_1(t_0)=-0.2915356957$.}
{0.45}{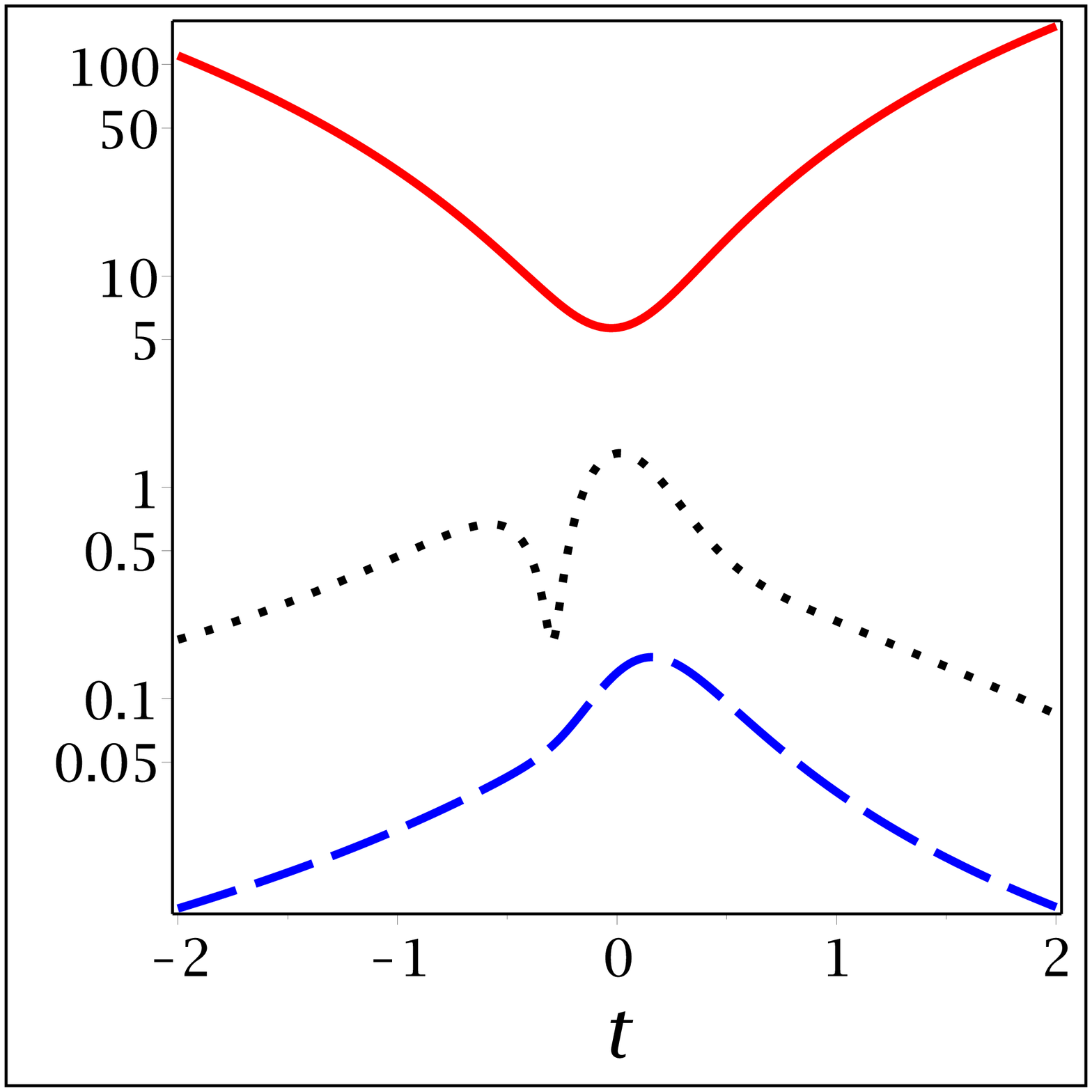}{0.45}{Quantum case for $c_0=-1$, ${\cal I}=0.5$,
$R=0.2$, $\alpha=+1$ and $c_1(t_0)=- 0.4454534878$.}{0.45}

\myfigures{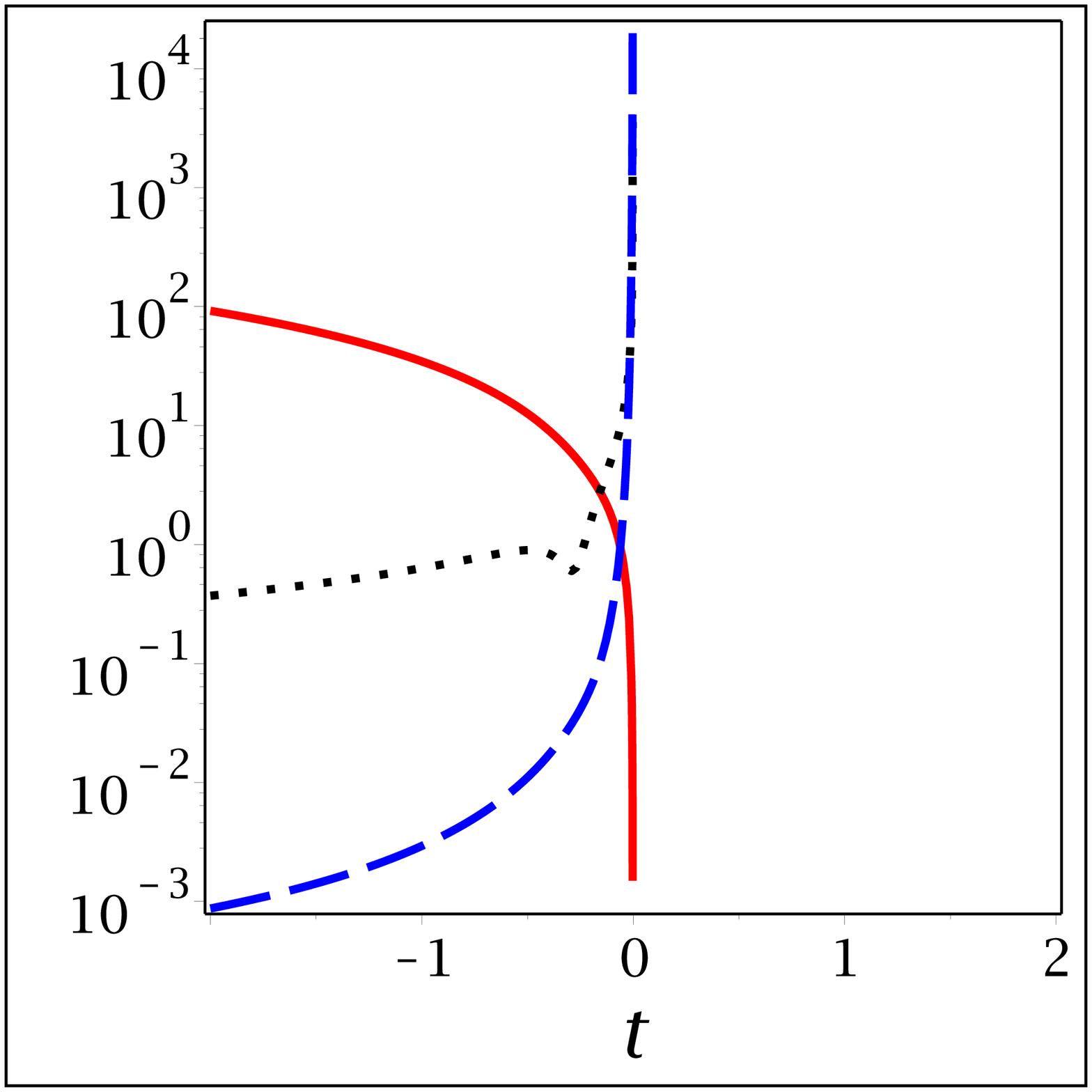}{0.45}{ Classical case for $c_0=-1$, ${\cal I}=0.5$,
$R=0.2$, $\alpha=-1$ and $c_1(t_0)=-0.1781242009$.}
{0.45}{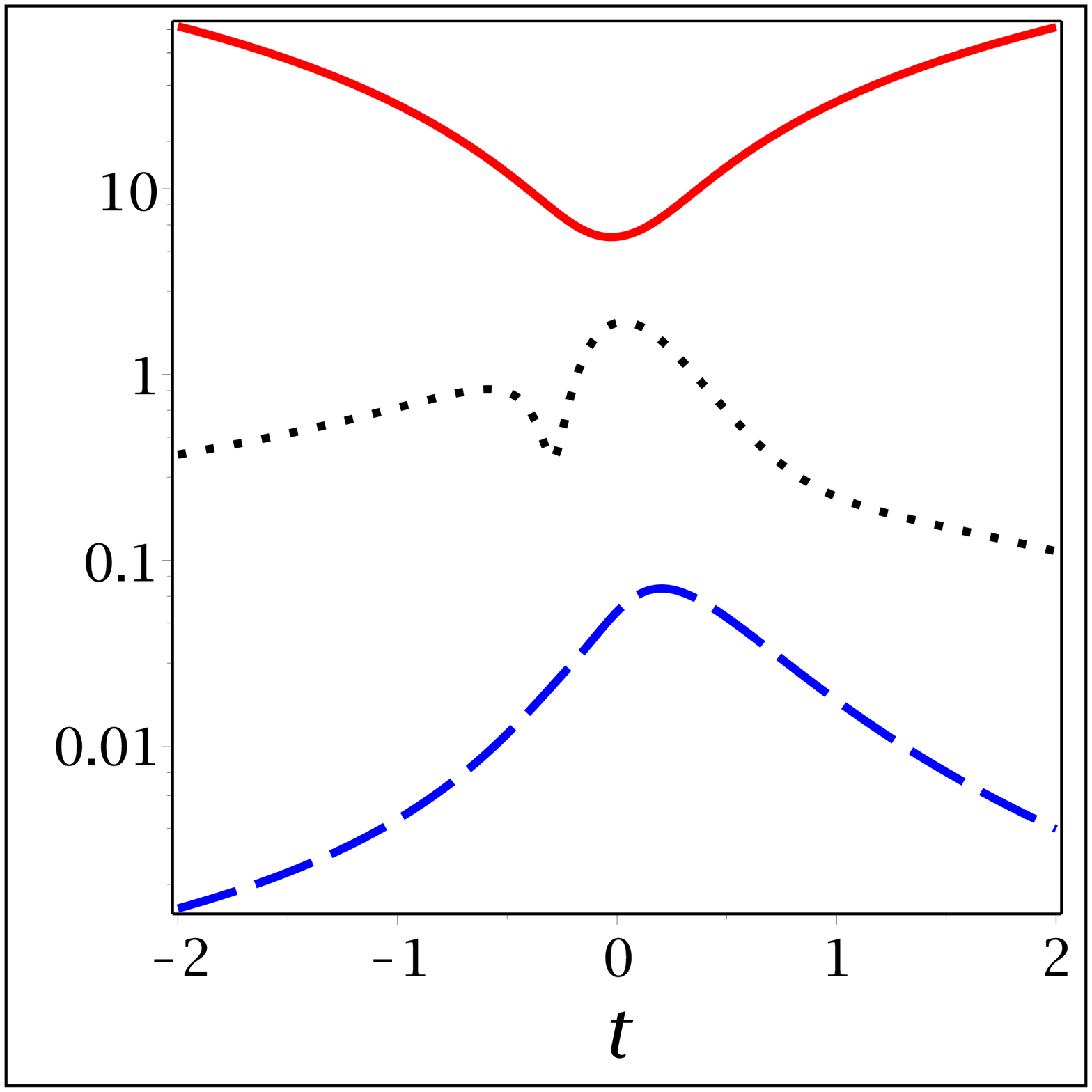}{0.45}{Quantum case for $c_0=-1$, ${\cal I}=0.5$,
$R=0.2$, $\alpha=-1$ and $c_1(t_0)=- 0.2993825979$.}{0.45}

\myfigures{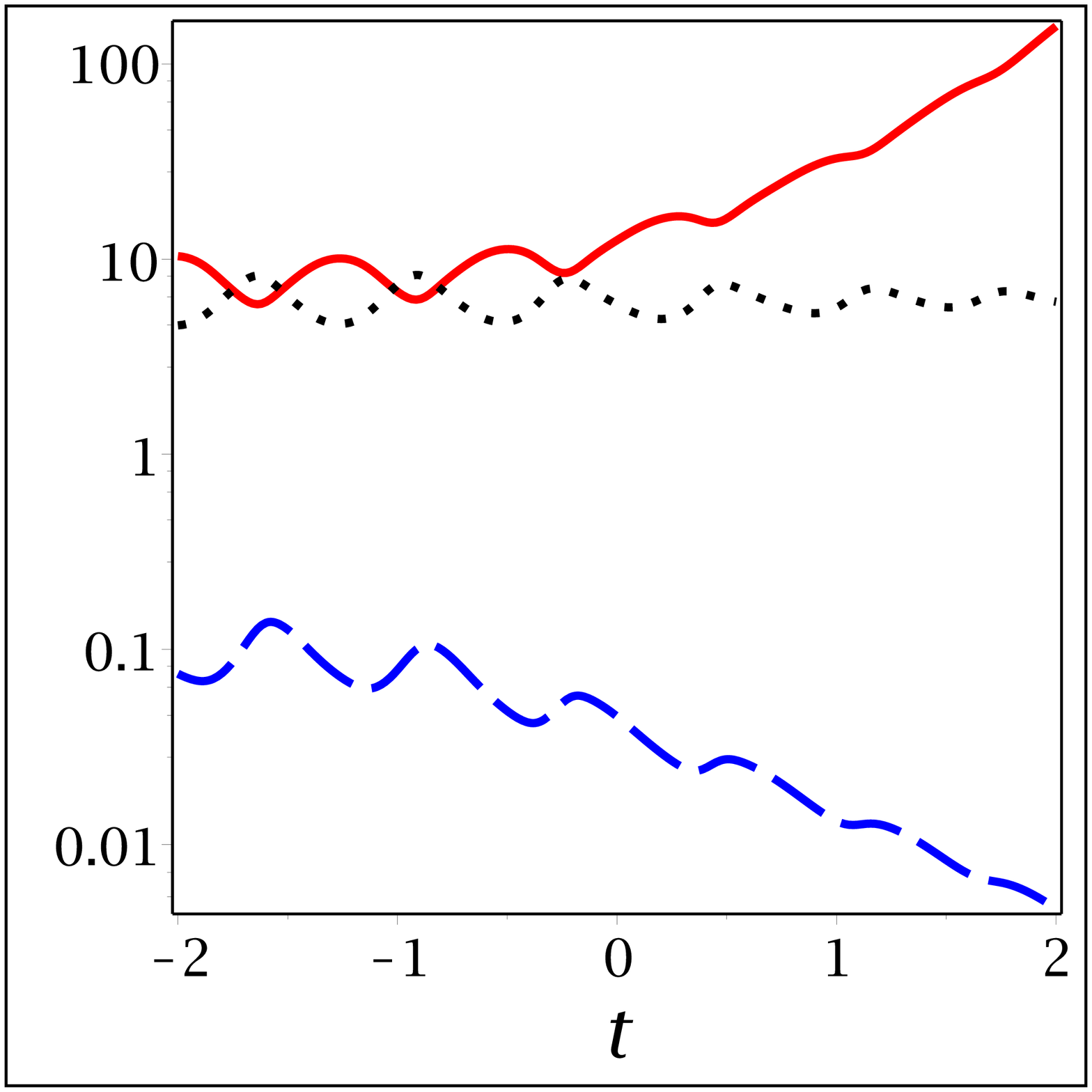}{0.45}{ Quantum case for $c_0=4$, ${\cal I}=0.5$,
$R=0.2$, $\alpha=+1$ and $c_1(t_0)=- 0.4993069119$.}
{0.45}{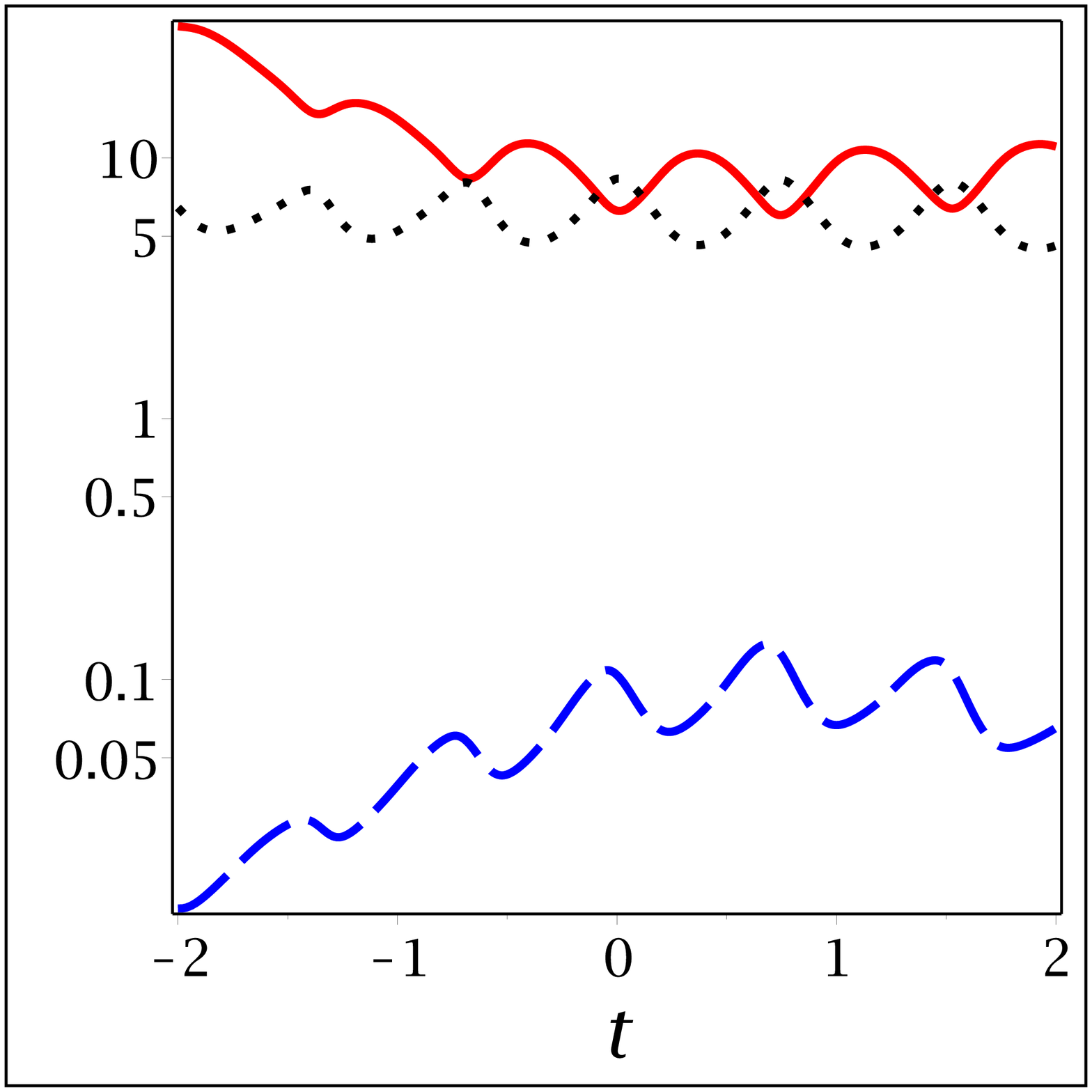}{0.45}{Quantum case for $c_0=-1$, ${\cal I}=0.5$,
$R=0.2$, $\alpha=+1$ and $c_1(t_0)=3.535164226$.}{0.45}

\myfigures{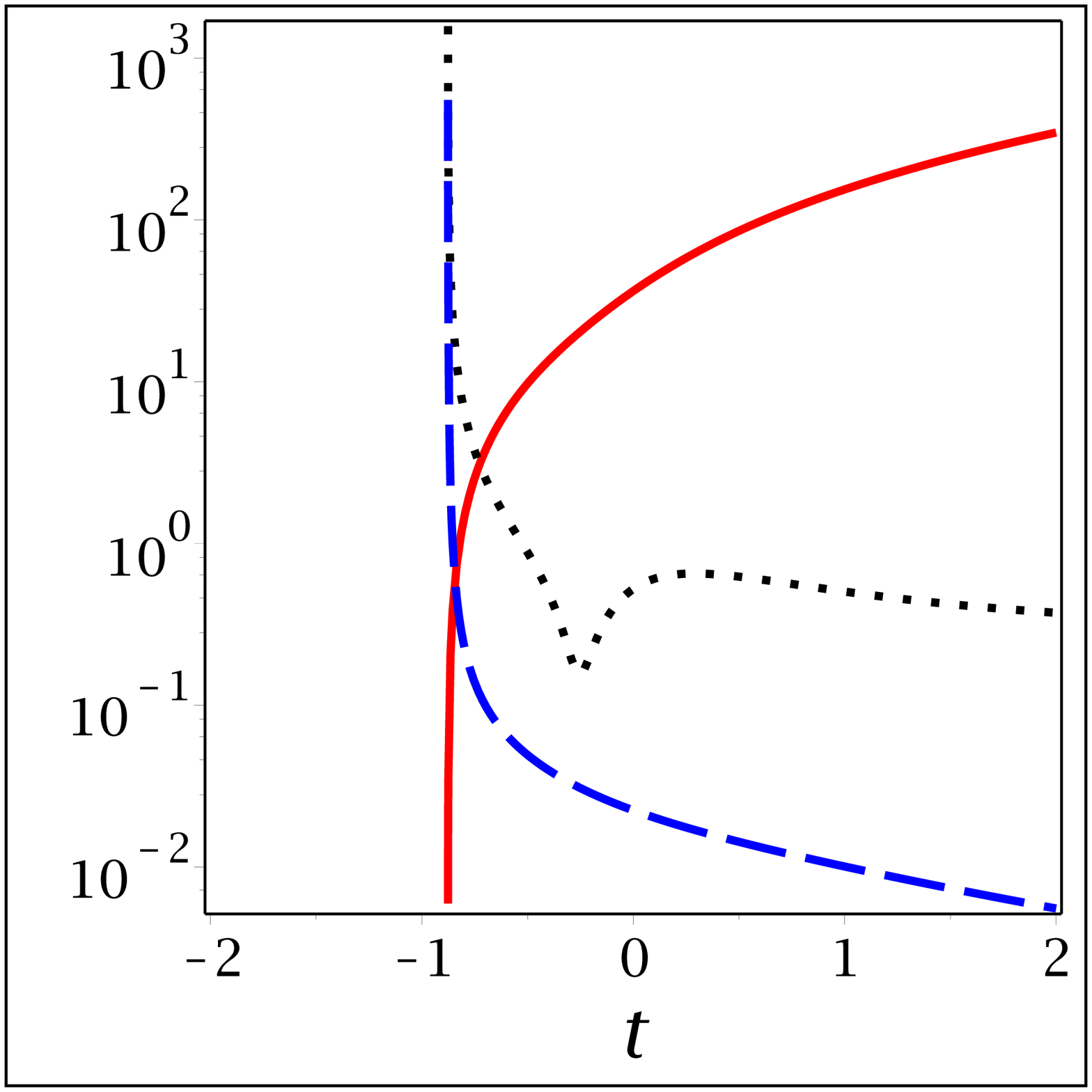}{0.45}{ Classical case for $c_0=4$, ${\cal I}=0.5$,
$R=0.2$, $\alpha=+1$ and $c_1(t_0)=0.8576784788$.}
{0.45}{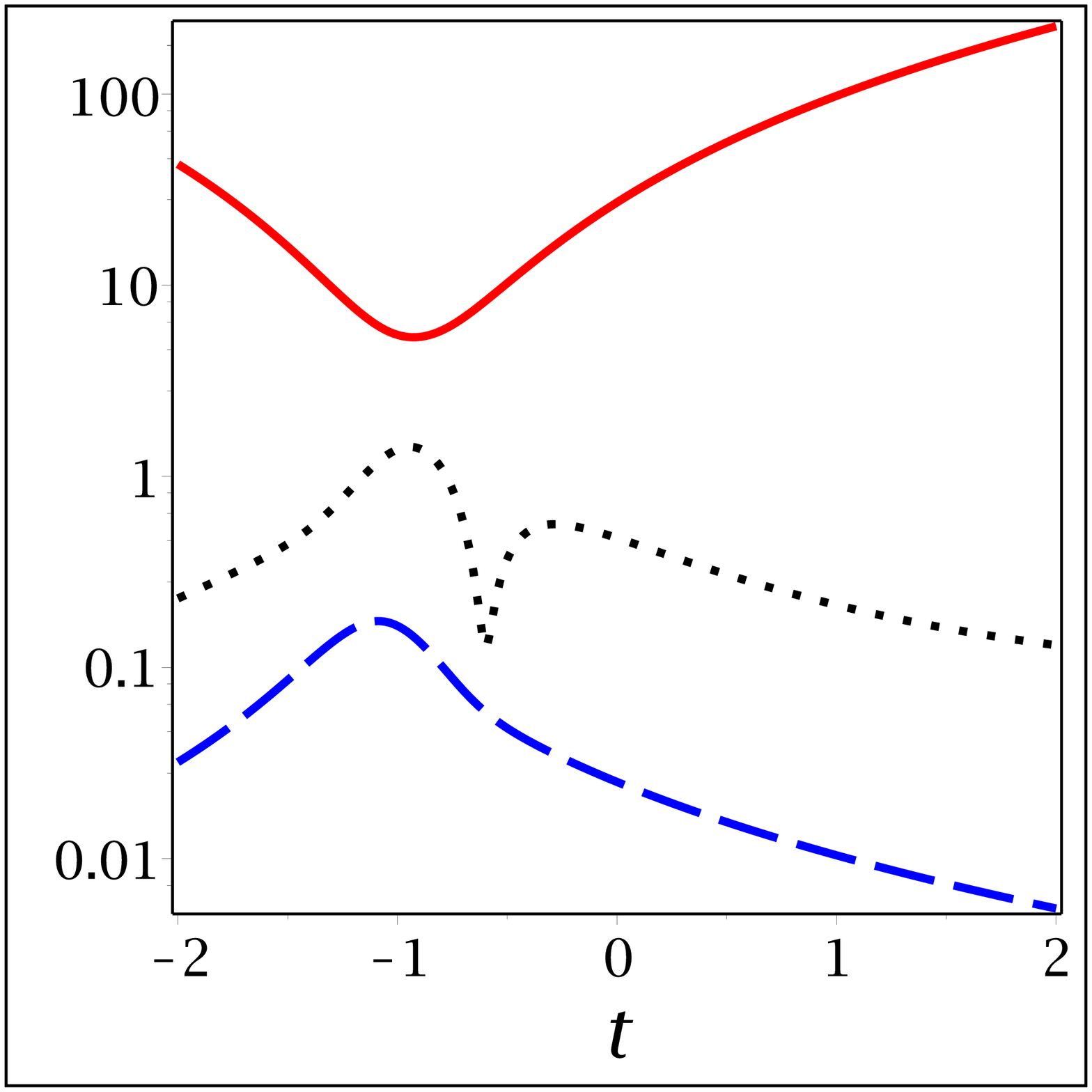}{0.45}{Quantum case for $c_0=4$, ${\cal I}=0.5$,
$R=0.2$, $\alpha=+1$ and $c_1(t_0)=3.589017650$.}{0.45}

In Figs. \ref{2103} and \ref{1113} we present the effective quantum
evolution of the BII universe with the initial conditions $c_0$ and
$c_1(t_0)$ of different signs. We remark the appearance of a few
oscillations in the vicinity of the classical singularity and from
which the solutions evolve smooth in time.

In Fig. \ref{5103} we present the evolution of the classical
evolution of volume scale $V$, matter density $\rho_{M}$ and shear
$\Sigma^2$ choosing all initial conditions $c_i(t_0) > 0$. THeir
quantum counterpart is plotted in Fig. \ref{2113}.  Again we get
that the classical BII universe evolves from a singularity, while in
the LQC approach all observable are finite. In the vicinity of the
classical singularity the effective solution presents a bounce and
subsequently the evolution is similar with the classical one.

%%%%%%%%%%%%%%%%%%%%%%%%%%%%%%%%%%%%%%%%%%

\section{Conclusions}

In this paper we analyzed the numerical solutions of the effective
equations of the LQC dynamics for BII model. We have extended the
effective LQC treatment of BII cosmologies by including cosmic
strings and a homogeneous magnetic field.

We considered the analytical and numerical solutions of BII model at
the classical and effective level. The main objective of the paper was
the investigation how the classical big bang singularity is resolved
and how the effective equations evolve. For this purpose we chose as
a set of observable quantities like volume, string density, shear
expansion and studied their evolution numerically. We showed that a big
bounce occurs in a collapsing magnetized BII string universe, thus
extending the known cosmological models of singularity avoidance. After
the bounce the universe enters a classical regime.

The numerical simulations are quite sensitive to the initial conditions
regarding the directional Hubble parameters. Choosing a negative $c_0$,
the numerical simulations are appropriate to describe the evolution
towards the big bang singularity. In the classical case the evolution
stops at the singularity, while in the quantum case we have bounces and
the singularity is eluded. On the other hand the choice of the initial
conditions with a positive $c_0$ is adequate to have in view the
evolution after the big bang. In the classical case, starting from the
vicinity of the singularity, the volume of the universe tends to
infinity, while the string density is smaller and smaller. In the
quantum case, after a few small bounces we have for large time the same
behavior as in the classical case.

As a final comment regarding the numerical simulations we note that
they are stable in respect of reasonable variations of the parameters
describing the cosmic strings.

The study of anisotropic models with different kinds of matter sources
in the framework of LQC deserves further investigations. These studies
will contribute to answer the challenge question if the bouncing
non-singular behavior of the effective solutions is generic.

%%%%%%%%%%%%%%%%%%%%%%%%%%%%%%%%%%%%%%%
\subsection*{Acknowledgments}

This work is supported in part by a joint Romanian-LIT, JINR, Dubna
Research Project, theme no. 05-6-1060-2005/2013.
M.V. is partially supported by program  PN-II-ID-PCE-2011-3-0137,
Romania.
%%%%%%%%%%%%%%%%%%%%%%%%%%%%%%%%%%%%%%%%%%
%%%%%%%%%%%%%%%%%%%%%%%%%%%%%%%%%%%%%%%%%%%

\end{document}